\documentclass[twocolumn]{aastex63}

\newcommand{\co}{${\rm^{12}CO}\,$} 	
\newcommand{\tco}{${\rm^{13}CO}\,$} 	
\newcommand{\eco}{${\rm C^{18}O}\,$} 	

\accepted{April 23, 2020}
\submitjournal{APJ}

\shorttitle{W40 and Serpens}
\shortauthors{T.S. et al.}

\begin{document}


\title{A detailed analysis on the cloud structure and dynamics in Aquila Rift}

\correspondingauthor{Tomomi Shimoikura}
\email{ikura@otsuma.ac.jp}

\author{Tomomi Shimoikura}
\affiliation{Otsuma Women's University \\
` Chiyoda-ku,Tokyo, 102-8357, Japan}

\author{Kazuhito Dobashi}
\affiliation{Tokyo Gakugei University \\
Koganei, 184-8501, Tokyo}

\author{Yoshiko Hatano}
\affiliation{Tokyo Gakugei University \\
Koganei, 184-8501, Tokyo}

\author{Fumitaka Nakamura}
\affiliation{National Astronomical Observatory of Japan, Mitaka, Tokyo 181-8588, Japan}
\affiliation{Department of Astronomical Science, School of Physical Science, SOKENDAI (The Graduate University for Advanced Studies), Osawa, Mitaka, Tokyo 181-8588, Japan}





\begin{abstract}

We present maps in several molecular emission lines of a 1 square-degree region covering the W40 and Serpens South molecular clouds belonging to the Aquila Rift complex. 
The observations were made with the 45 m telescope at the Nobeyama Radio Observatory.
We found that the $^{12}$CO and $^{13}$CO emission lines consist of several velocity components with different spatial distributions.
The component that forms the main cloud of W40 and Serpens South, which we call the ``main component",
has a velocity of $V_{\rm LSR}\simeq 7$ km s$^{-1}$.
There is another significant component at $V_{\rm LSR}\simeq 40$ km s$^{-1}$,
which we call the ``40 km s$^{-1}$ component".
The latter component is mainly distributed around two young clusters: W40 and Serpens South. 
Moreover, 
the two components look spatially anti-correlated. 
Such spatial configuration suggests that the star formation in W40 and Serpens South was induced by the collision of the two components.
We also discuss a possibility that the 40 km s$^{-1}$ component consists of gas swept up by 
superbubbles created by SNRs and stellar winds from the Scorpius-Centaurus Association.
\end{abstract}






\section{Introduction} \label{sec:intro}

\begin{figure*}
\begin{center}
\includegraphics[scale=0.4]{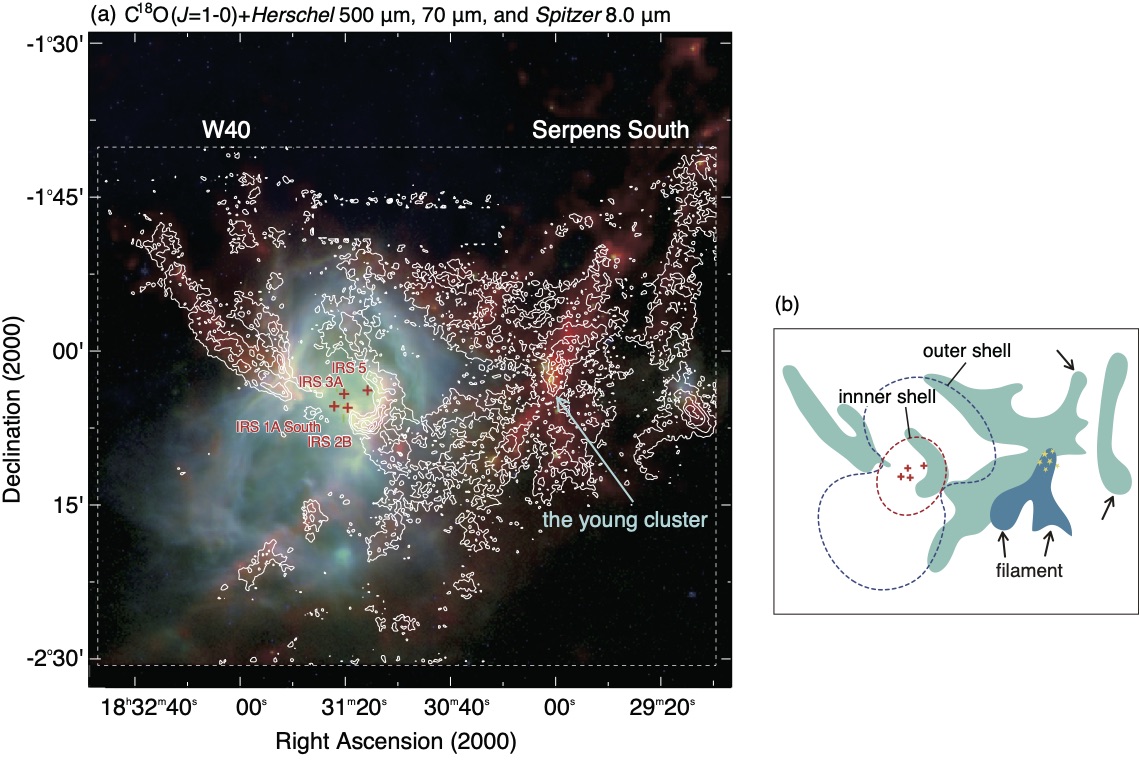}
\caption{
(a)Overview of the structure of the W40 and Serpens South regions. 
The contours represent the integrated intensity map of the \eco emission line (taken from Figure 3 of  Shimoikura et al. 2018), which is overlaid on the $\it{Herschel}$ $500 \micron$ (red), $70 \micron$ (green) and $\it{Spitzer}$ $8.0 \micron$ (blue) composite infrared image. 
The plus signs indicate the positions of the high mass infrared sources IRS 1A South, IRS 2B, IRS 3A, and IRS 5 \cite[e.g.,][]{Smith, Shuping2012}.
The regions enclosed by the dashed line indicates the observed area using the C$^{18}$O emission line.
(b) Schematic illustration of the clouds, the filaments, and the two shells in the W40 and Serpens South regions constructed from the C$^{18}$O observations \citep{Shimoikura2018}.
\label{fig:C18O}}
\end{center}
\end{figure*}

W40 and Serpens South belong to the Aquila Rift complex, located at $\ell \simeq28.8\arcdeg$ and $b \simeq3.5\arcdeg$.
Hundreds of young stellar objects (YSOs) have been identified in both regions \citep[e.g.,][]{Kuhn, Maury2011, Konyves2015}.
In a $\sim11$ square degrees field around the regions, 
\cite{Konyves2015} identified 651 starless cores most of which are gravitationally bound prestellar cores through the $\it{Herschel}$ observations. 
They also identified more than 400 prestellar cores and estimated the star formation efficiency within an individual prestellar core to be $\sim40\%$.
The presence of many molecular outflows in both regions also has been confirmed \citep[e.g.,][]{Nakamura2011,Shimoikura2015}.
W40 and Serpens South are one of the most active star-forming regions in the Galaxy.
Figure \ref{fig:C18O}(a) shows an overview of the structure of the W40 and Serpens South regions,
along with the distributions of C$^{18}$O emission revealed by our previous studies \citep{Shimoikura2018}.

W40 is an H{$\,${\sc ii}} region, also known as Sh-2 64 \citep{Sharpless}. 
There are high mass infrared sources in the central region (i.e., IRS 1A South, IRS 2B, IRS 3A, and IRS 5) that illuminate the H{$\,${\sc ii}} region \citep{Smith, Shuping2012}.
A dense dark cloud with a diameter of about $20\arcmin$ surrounds the H{$\,${\sc ii}} region \citep{Vallee,Dobashi2005, Dobashi2011}. 
The ages of the YSOs forming in the cloud are estimated to be $\sim1$ Myr \citep[e.g.,][]{Kuhn}.

An infrared dark cloud designated Serpens South lies just west of W40. 
Serpens South is accompanied by an extremely young cluster \citep[$\sim0.5~$Myr old,][]{Gutermuth2008}.
It is composed of several filaments of gas and dust.

\cite{Nakamura2014} identified six dense ridges in the filaments that appear to be colliding to form the young cluster.
Near-infrared polarization observations have revealed that the filaments are penetrated by globally-ordered magnetic fields.
The global magnetic field is perpendicular to one of the filaments containing the young cluster, suggesting that the magnetic field plays a crucial role in the cloud evolution \citep{Sugitani2010,Kusune2019}.

Previous observations using various molecular lines have shown that the two clouds have complex velocity structures \citep{Zhu, Pirogov2013, Nakamura2014, Shimoikura2015, Shimoikura2018},
and they have the same velocity at $V_{\rm LSR}\simeq 7$ km s$^{-1}$. 
In our previous paper \citep{Shimoikura2018}, 
we examined the spatial distributions of the clouds.
Results obtained using the C$^{18}$O molecular emission line show that the spatial and velocity structures are distributed continuously across the W40 and Serpens South regions and that they are physically connected \citep{Shimoikura2018}.
We also found that two elliptical structures centered on the H{$\,${\sc ii}} region in the position-velocity diagram, 
indicating that they can be explained as parts of two expanding shells:
one is an inner shell of $\sim0.5$ pc in diameter centered on the H{$\,${\sc ii}} region,
and the other is an outer shell of $\sim2.5$ pc in diameter 
that corresponds to a boundary of the H{$\,${\sc ii}} region.
The dense gas associated with the young cluster in Serpens South is likely to be located in the outer shell, suggesting that the expansion of the shell induced the cluster formation there \citep{Shimoikura2018}.
In Figure \ref{fig:C18O}(b), we show the schematic illustration of the clouds, the filaments, and the two shells in the W40 and Serpens South regions constructed from the previous results.

It has been suggested that the stellar winds due to the star formation in the Scorpius-Centaurus Association (Sco OB2) caused sequential star formation and that their influence may also affect the entire Aquila Rift, to which W40 and Serpens South belong \citep{deGeus1992,Frisch1995,Preibisch2008}.
Star formation in Sco OB2 created several large-scale interstellar H{$\,${\sc i}} shells that provide the basis for linking nearby bubble-like structures such as Loop I \citep[e.g.,][]{Frisch2018}.
From large-scale CO mapping observations of about 10 square degrees around W40 and Serpens South, 
\cite{Nakamura2017} found that the surroundings of W40 and Serpens South also show complex spatial and velocity structures,
and they identified multiple large-scale velocity-coherent structures extending over $\sim10$ pc, which they called ``arcs". 
They concluded that the arcs might be created by bubbles 
converging toward W40 and Serpens South.
They also speculated that the interaction of the molecular clouds with the bubbles
may have triggered star formation in W40 and Serpens South. 
However, the relationship between star formation in W40 and Serpens South with the surrounding gas is still incompletely understood.

In this paper, we present maps of the $^{12}$CO and $^{13}$CO emission lines covering an area of 1 square degree around W40 and Serpens South with an effective beam size of $\sim20\arcsec$.
In a previous study \citep{Nakamura2019, Shimoikura2018}, in addition to the velocity component at $V_{\rm LSR}\sim 7$ km s$^{-1}$ that contains the W40 and Serpens South clouds, 
we pointed out the existence of multiple independent spatial and velocity components.
However, no investigation has been conducted on the relationship between those components and W40 and Serpens South.
We here investigate the relationship between the gas dynamics around W40 and Serpens South and star formation there.

Estimates of the distance to W40 and Serpens South vary from 250 to 600 pc \cite[e.g.,][]{Rodney}. 
Recently, \cite{Ortiz2018} obtained consistent results between $\it{Gaia}$ DR2 data and the VLBA measurements for the mean parallax in the regions, and they concluded that W40 and Serpens South are located at a distance of $436 \pm 9$ pc.
We adopt the distance 436 pc obtained by \cite{Ortiz2018} for the observed area, which we also used in \cite{Shimoikura2018}.

This paper is structured as follows.
In Section 2, we present the observational procedures. 
The results of the gas distributions around the observed region are presented in Section 3;
we found several velocity components and estimated their masses.
In Section 4, we discuss the relationship between W40 and Serpens South.
Finally, we summarize our results in Section 5.


\section{OBSERVATIONS} \label{sec:obs}

The observations reported in this paper were conducted using the 45 m telescope at the Nobeyama Radio Observatory (NRO) 
as part of ``the Star Formation legacy project" \citep{Nakamura2019}.
We carried out mapping observations toward the W40 and Serpens South regions covering an area of $1\arcdeg\times1\arcdeg$ 
using $5\arcsec$ spacing and 0.05 km s$^{-1}$ velocity resolution in the five emission lines, i.e., \co (115.271204 GHz),
\tco (110.201354 GHz), \eco (109.782176 GHz), CCS (93.870098 GHz), and N$_2$H$^+$ (93.1737637 GHz).
The Half Power Beam Width (HPBW) of the telescope is $\sim15\arcsec$ at 115 GHz.
We used the FOur-beam-REceiver System on the 45 m Telescope \citep[`FOREST' ;][]{Minamidani} as the frontend and the digital spectrometer `SAM45' as the backend.
Typical system temperatures were $\sim300$ K for the \co line and $\sim250$  K for the \tco line. 
To check the telescope pointing, 
we observed the SiO maser source V1111-Oph every 1.5 hr
and found that the accuracy was within $\sim5\arcsec$.

We performed the data reduction using NOSTAR \citep{Sawada}, 
a reduction software package developed at NRO.
We converted the antenna temperature $T_{\rm a}$ to the main beam brightness temperature $T_{\rm mb}$ 
using the main beam efficiency $\eta_{\rm mb}\simeq 0.4-0.5$ which changes depending on the frequencies. 
We finally obtained spectral data at an effective angular resolution of $22\arcsec-24\arcsec$ 
with a velocity resolution of 0.1 km s$^{-1}$.
The typical rms noise temperatures in $T_{\rm mb}$ are in the range $0.3-0.9$ K for the observed emission lines.
The spatial resolution $\sim22\arcsec$ corresponds to $\sim0.05$ pc in the observed region. 

A more detailed description of the observations is presented in our previous paper \citep{Nakamura2019,Shimoikura2018}.


\section{RESULTS}\label{sec:results}

\subsection{Spatial Distributions of the Two Componens}\label{sec:2comp}

\begin{figure*}
\begin{center}
\includegraphics[scale=0.4]{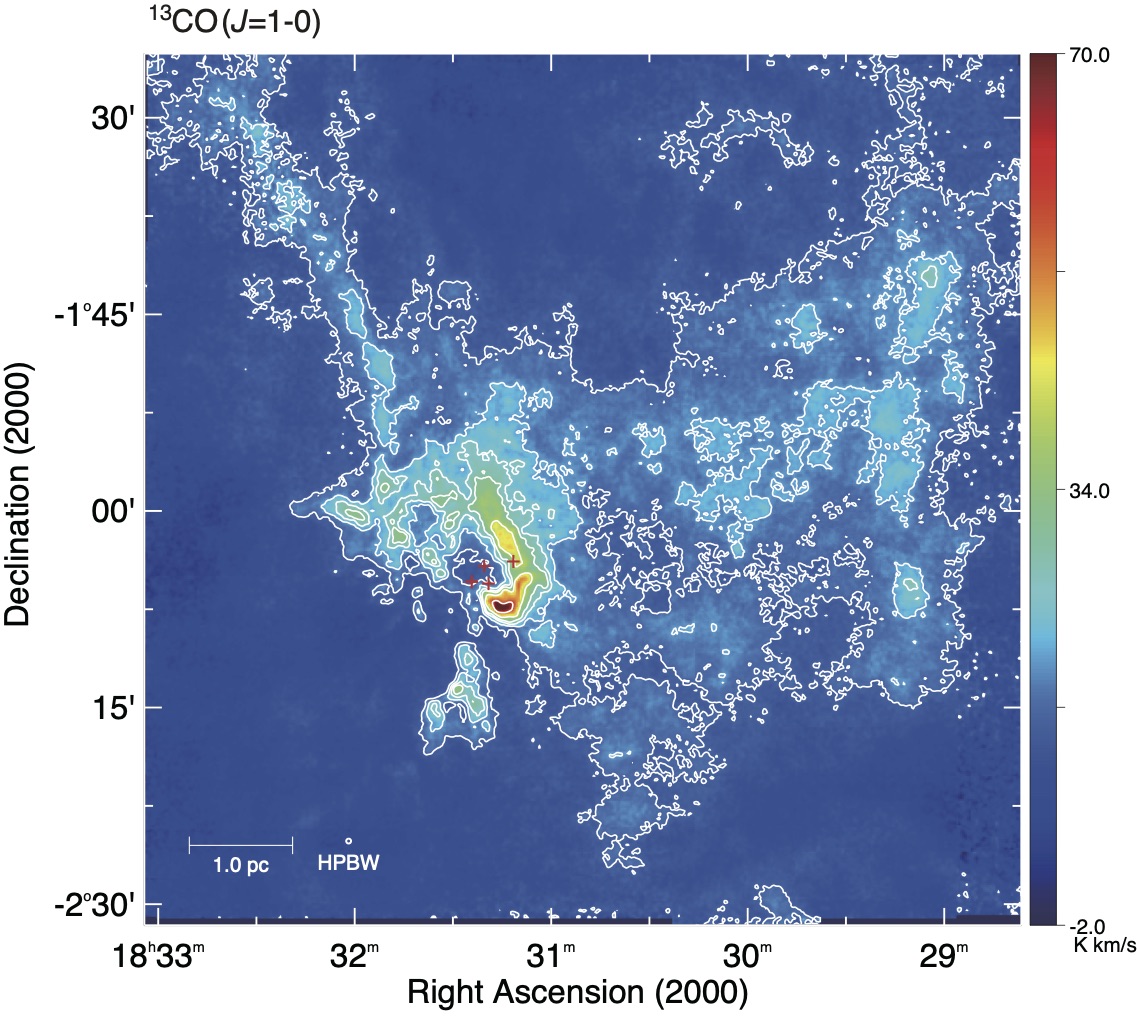}
\caption{
The \tco intensity map integrated over the velocity range of  0 to 15 km s$^{-1}$.
The contours are drawn at 15, 20, 25, 30, 40, and 60 K km s$^{-1}$.
\label{fig:13CO}}
\end{center}
\end{figure*}

Figure \ref{fig:13CO} shows the integrated intensity map of the \tco emission line. 
Within the two regions of W40 and Serpens South, the \eco emission is strongly detected in the vicinity of each cluster-forming region (Figure \ref{fig:C18O}), 
whereas the \tco emission shows a clear peak only around the cluster in W40 but appears more flattened around Serpens South probably because of the heavy absorption by colder gas in the foreground \citep{Shimoikura2015}.
Some of the filaments found in the $\it{Herschel}$ data \citep[e.g.,][]{Andre2010} have also been detected in the \tco emission line.

\begin{figure*}
\begin{center}
\includegraphics[scale=0.4]{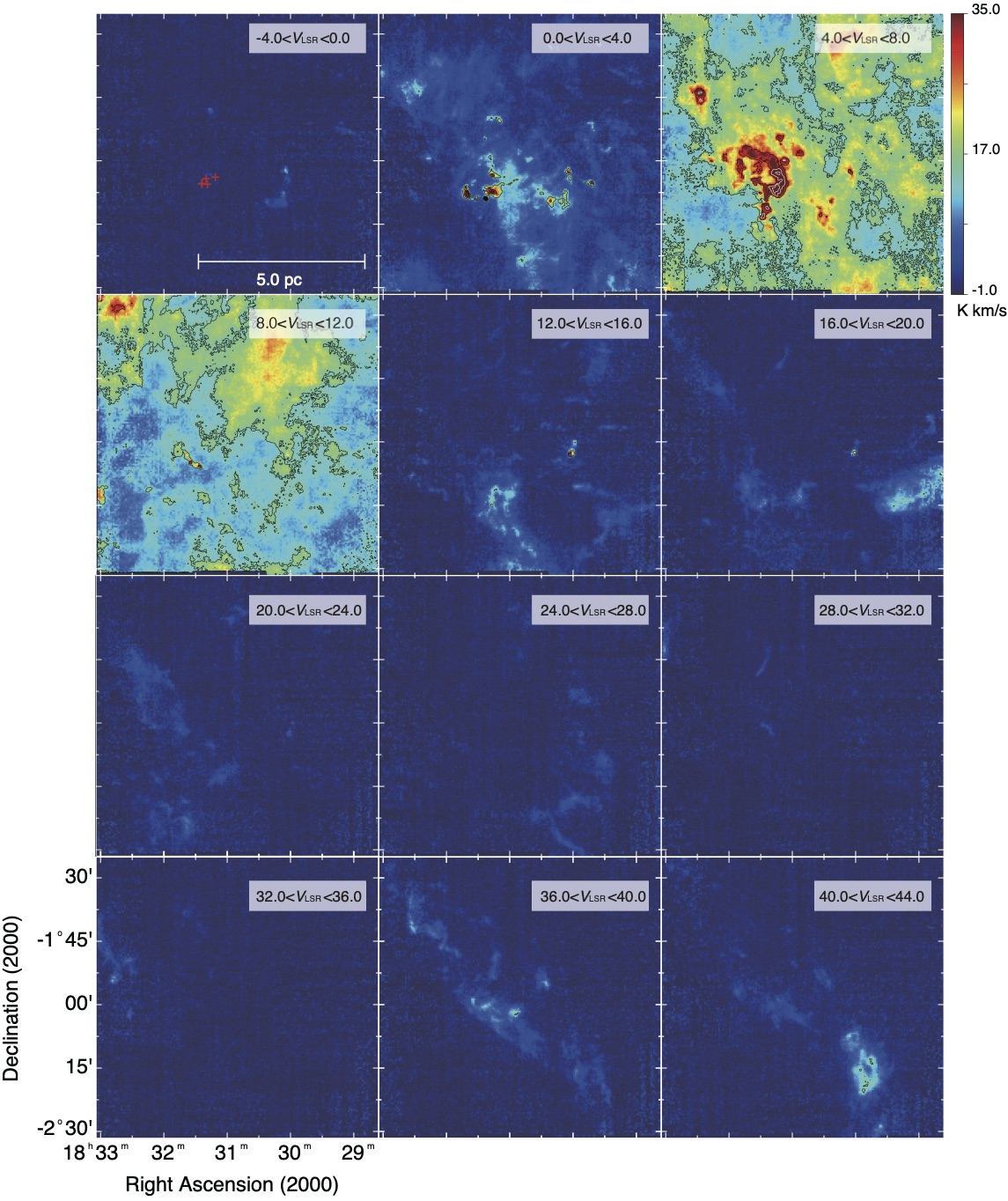}
\caption{
Velocity channel maps of the \co emission line drawn at every 4 km s$^{-1}$.
The contours are drawn from 15 K km s$^{-1}$ with a step of 15 K km s$^{-1}$. 
\label{fig:12CO_channel}}
\end{center}
\end{figure*}

\begin{figure*}
\begin{center}
\includegraphics[scale=0.4]{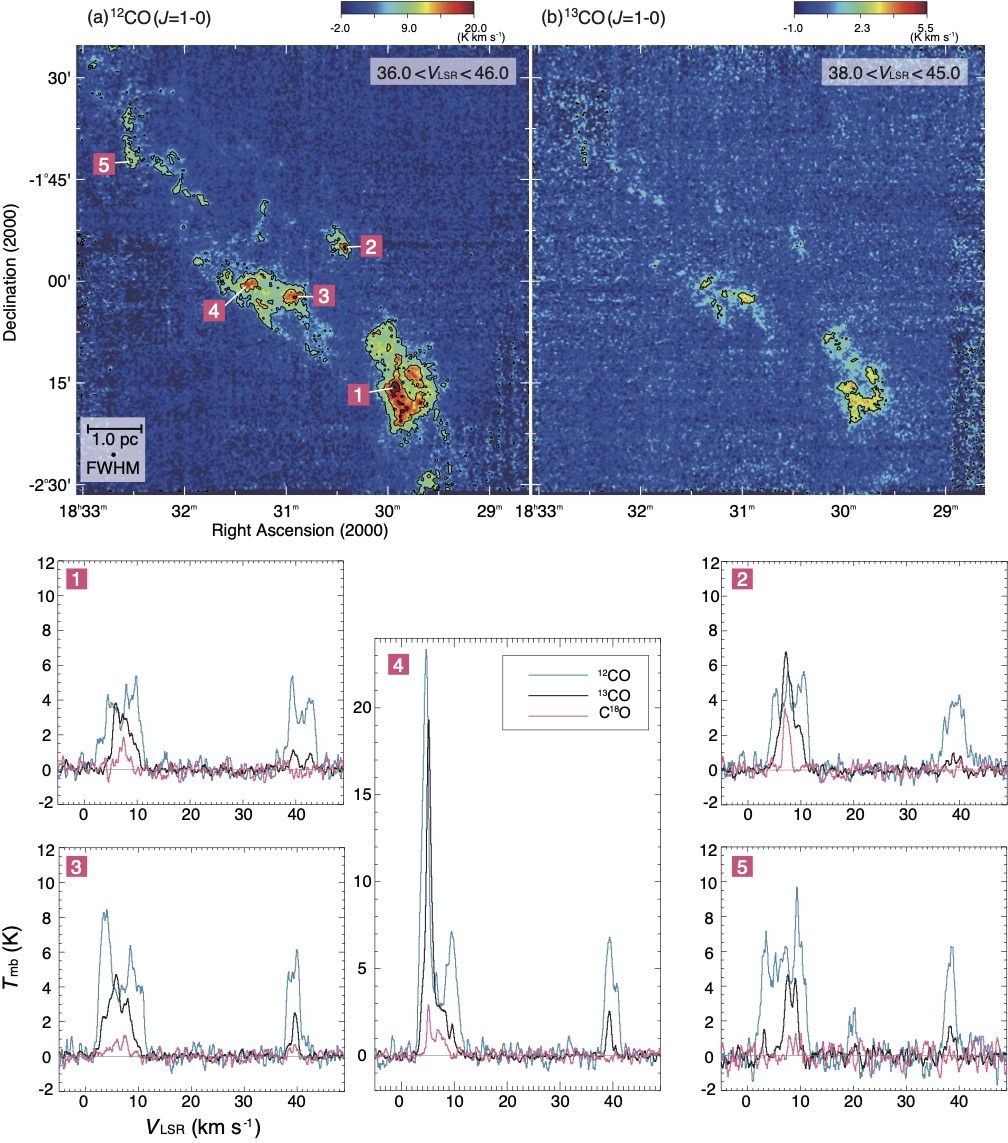}
\caption{
Distribution of the 40 km s$^{-1}$ component traced in (a) the \co emission line and (b) the \tco emission line.
The velocity range used for the integration, in units of km s$^{-1}$, is indicated in the top right corner of each panel. 
The lowest contours and the contour intervals are 6.0 K km s$^{-1}$ for the \co map, and 2.2 K km s$^{-1}$ for the \tco map. 
The \co (blue line), \tco (black line) and \eco (red line) spectra taken at the positions in panel (a) are displayed in the correspondingly numbered panels. 
\label{fig:spectra}}
\end{center}
\end{figure*}

Figure \ref{fig:12CO_channel} shows the velocity channel maps of the \co emission line made every 4 km s$^{-1}$ 
in the velocity range $-4\leqq V_{\rm LSR}\leqq44$ km s$^{-1}$. 
The \co emission line is strongly detected in the range $0\leqq V_{\rm LSR}\leqq12$ km s$^{-1}$. 
The \tco emission line is also strongly detected in the almost same velocity range. 
We call the velocity component detected in this velocity range the ``main component". 
In addition to the main component, there are some other independent velocity components, at $V_{\rm LSR} \simeq 20, \simeq 25, \simeq 35$, and $\simeq 40$ km s $^{-1}$, in the observed area.
Other than the main component, 
we found that the velocity component around $40$ km s$^{-1}$ has the highest intensity and is distributed over a wide area. 
We call this component the ``40 km s$^{-1}$ component".
In Figure \ref{fig:spectra}, we show the \co and \tco distributions of the 40 km s$^{-1}$ component 
as well as the spectra of the three CO emission lines sampled at several positions.
We found that the 40 km s$^{-1}$ component is also detected in C$^{18}$O in panel 3 of spectrum though the emission is weak ($T_{\rm mb}\simeq 1$ K).

\begin{figure*}
\begin{center}
\includegraphics[scale=0.4]{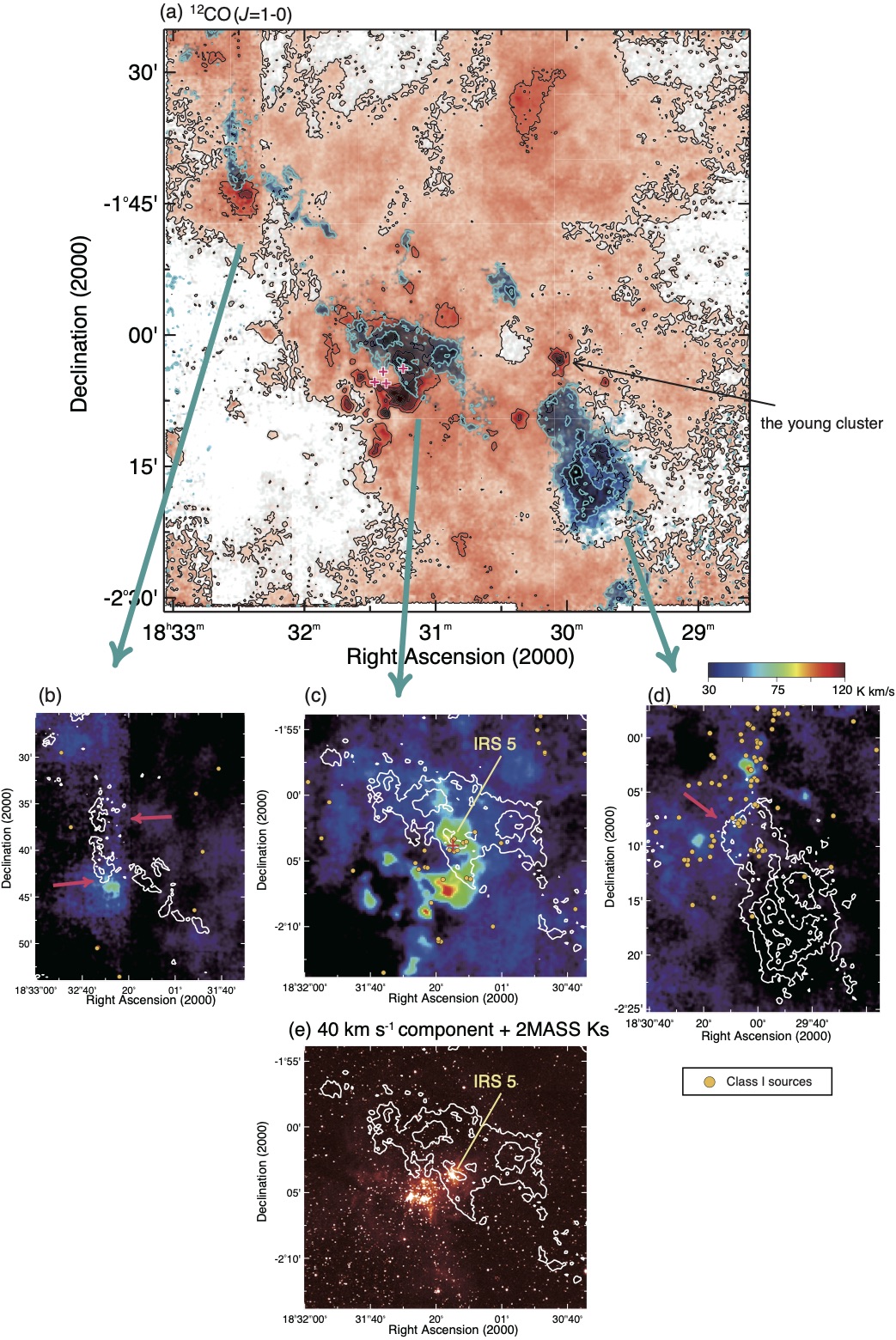}
\caption{
(a) Distribution of the main component (red scale) and the 40 km s$^{-1}$ component (blue scale) traced in the \co emission line.
For the main component, the integrated velocity range is $V_{\rm LSR}=0-15$ km s$^{-1}$.
The lowest contour is 30.0 K km s$^{-1}$, and the contour interval is 20.0 K km s$^{-1}$.
For the 40 km s$^{-1}$ component, the integrated velocity range and the contours are the same as in Figure \ref{fig:spectra}(a).
Panels (b)--(d) show the close-up views around the area indicated by the green arrow.
The main component is shown by the color scale and the 40 km s$^{-1}$ component is shown by the contours.
The yellow circles denote the Class I sources identified in the WISE point source catalog using the selection
criteria suggested by \cite{Koenig}.
(e) The 40 km s$^{-1}$ component superposed on the 2MASS Ks band image.
\label{fig:2comp}}
\end{center}
\end{figure*}

We compare the distributions of the main and 40 km s$^{-1}$ components traced in \co in Figure \ref{fig:2comp}(a).
The 40 km s$^{-1}$ component is distributed mainly around the H{$\,${\sc ii}} region 
and vicinity of the young cluster in Serpens South.
It is also seen in the northeast part of the observed area.
The scale-up views of these parts are shown in panels (b)-(e) of the figure.
In panels (b) and (d), 
the spatial distribution of the main component seems to be anti-correlated with that of the 40 km s$^{-1}$ component as indicated by the red arrows. 
In panel (c), 
we found that the spatial distributions of the two components overlap each other.
As shown in panel (e), in W40, there is a compact H{$\,${\sc ii}} region around the B-type star IRS 5 \citep{Mallick}, which is an evidence for star formation in progress.
In order to investigate spatial distributions of the young stellar objects associated with the components, we used the WISE point source catalog in the observed regions.

In panels (b)--(d), we plot the selected Class I sources identified in the catalog using the criteria suggested by \cite{Koenig}, which is based on the WISE color-color classification technique.
In panel (c), we found that some Class I sources centered on IRS 5 are located where the two components overlap.

\begin{figure*}
\begin{center}
\includegraphics[scale=0.4]{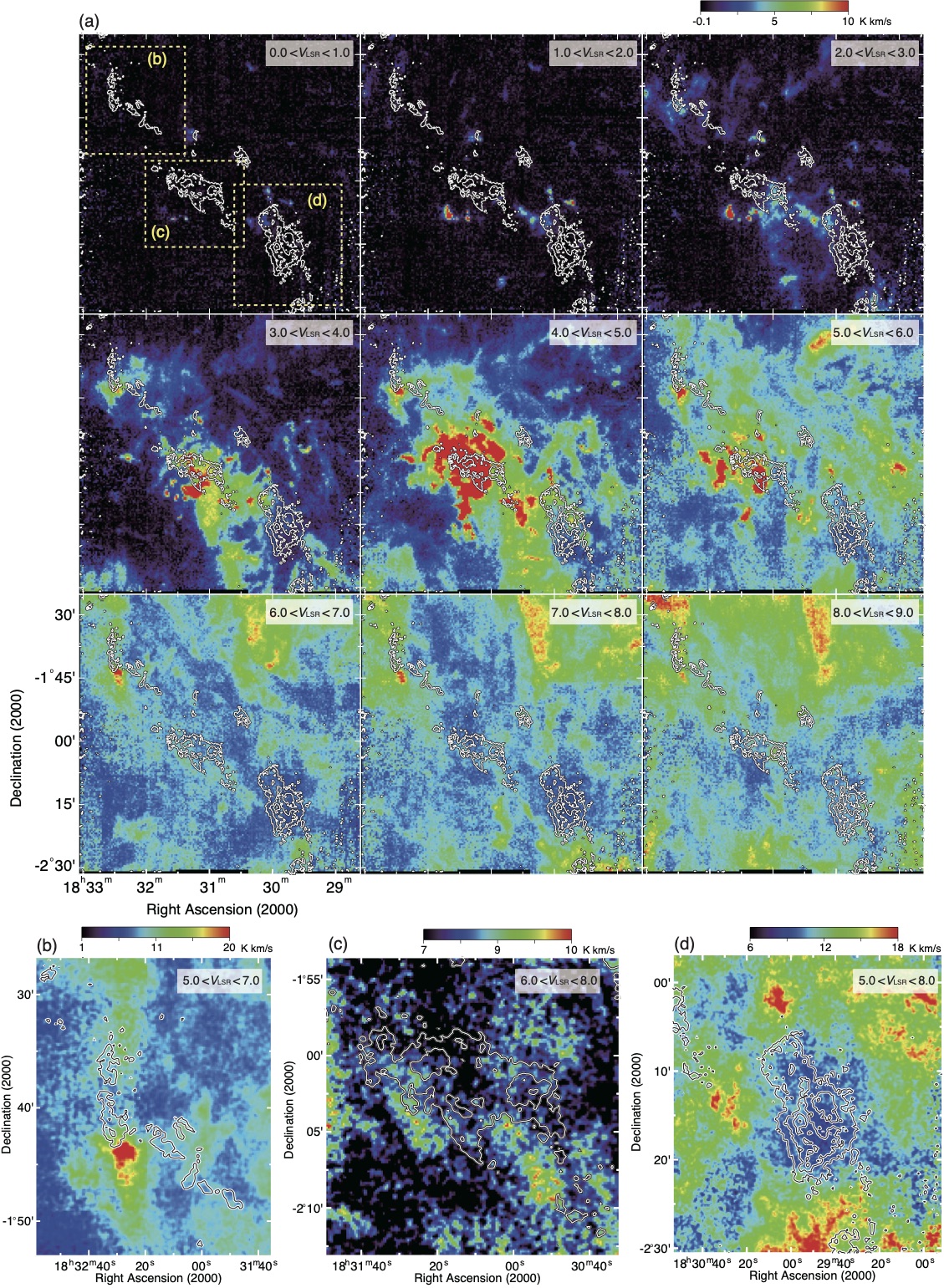}
\caption{
(a) Velocity channel map of the \co emission line in the velocity range $0\leqq V_{\rm LSR}\leqq 9$ km s$^{-1}$ superposed on the 40 km s$^{-1}$ component as in Figure \ref{fig:spectra}.
(b)--(d) Enlarged views toward the dashed regions in panel (a), 
but the color scale shows the \co intensity map integrated over the velocity range shown in each panel.
\label{fig:channel2}}
\end{center}
\end{figure*}

In order to examine the relationship between the two components more carefully,
we made the velocity channel maps of the main component with a finer velocity interval (1 km s$^{-1}$) than Figure \ref{fig:12CO_channel}
and compared them with the 40 km s$^{-1}$ component in Figure \ref{fig:channel2} (a).
In the three parts shown in Figure \ref{fig:2comp}, 
we found that there is an anti-correlation between the distribution of the main component in the specific velocity range and that of the 40 km s$^{-1}$ component, which are shown in panels (b)--(d) in Figure \ref{fig:channel2}.
There is a clear ``hole" of the main component and 
it roughly coincides with the distribution of the 40 km s$^{-1}$ component,
which is noticeable in panel (d).
These results suggest that the main and 40 km s$^{-1}$ components are physically interacting.



\subsection{Mass Estimates for the Two Components}\label{sec:mass}

Using the H$_2$ column density, $N$(H$_2$), from the archival data of the $\it{Herschel}$ Gould Belt survey \citep[cf.][]{Andre2010},
we estimate the total mass of the entire observed region to be
\begin{equation}
\label{eq:Mass}
M=\alpha m_\mathrm{H}S_\mathrm{pix} \Sigma N(\mathrm{H}_2)
\end{equation}
where $\alpha$ is the mean molecular weight taken to be 2.8, $m_{\rm H}$ is the mass of a hydrogen atom,
and $S_{\rm pix}$ is the pixel area ($7.5\arcsec\times7.5\arcsec$).
The mass obtained by integrating $N$(H$_2$) within the area defined by the $5\times10^{21}$ H$_2$
cm$^{-2}$ contour is estimated to be $1.2\times10^{4}M_{\sun}$.

The value of $N$(H$_2$) derived from the dust data obtained by $\it{Herschel}$ is affected by line-of-sight contamination, and 
it includes diffuse emission not related to the two components. 
Thus, we estimate the molecular mass of the two components 
from the \tco data, assuming local thermal equilibrium (LTE) 
and the optically thin limit.

We computed the optical depth of the \tco emission, $\tau{(^{13}\rm{CO})}$, at each position as 
\begin{equation}
\label{eq:Tau}
\tau{(^{13} \rm{CO})}=-\mathrm{ln}\left[1-\frac{T{\rm_{mb}^{^{13}\rm{CO}}}}{J(T_{\mathrm{ex}})-J(T\mathrm{_{bg}})}\right]
\end{equation}
where $J(T)=T_{0} \diagup(e^{(T_{0} \diagup T)}-1)$ with $T_{0}=h\nu/k$. 
$T_{\rm bg}$ is the cosmic background temperature 2.73 K.
For the excitation temperature $T_{\rm ex}$, we used the $T_{\rm ex}$ data from our previous study, where we determined it from the $T_{\rm mb}$ data of the optically thick \co line \cite[][see their Figure 6]{Shimoikura2018}.

The column density of \tco, $N$(\tco), is then given by
\begin{equation}
\label{eq:column1}
N(^{13}\mathrm{CO})=\frac{3h}{8\pi^{3}}\frac{Q}{\mu^{2}S_{ij}}\frac{e^{Eu/k{T}_{\mathrm{ex}}}}{e^{T_{0}/T_{\mathrm{ex}}}-1}{\Delta V \tau{(^{13} \rm{CO})}}{}
\end{equation}
where $Q$ is the partition function defined as $kT_{\rm ex}/hB_{0}$ where $B_{0}$ is the rotational constant of the molecule. 
$\mu$ is the dipole moment, $E_{u}$ is the energy of the upper level, and $S_{ij}$ is the intrinsic line strength of the transition from state $i$ to $j$. 
The spectral line parameters are taken from Splatalogue\footnote{www.splatalogue.net}. 
$\Delta V$ is the observed line width of the $^{13}$CO emission.
We applied a Gaussian fitting to the $^{13}$CO emission line at all of the observed positions to measure $T{\rm_{mb}^{^{13}\rm{CO}}}$ and $\Delta V$. 
We employed an $N$(H$_2$)/$N$(\tco) conversion factor of $5.0\times10^{5}$\citep{Dickman1978}.
Here, among the observed area, there are positions where the $^{12}$CO emission line is lower than $T{\rm_{mb}^{^{13}\rm{CO}}}$
due to self-absorption in the $^{12}$CO emission line around at $\sim7$ km s$^{-1}$ 
(e.g., see panel 1 of Figure \ref{fig:spectra}).
For such positions, $\tau{(^{13}\rm{CO})}$ cannot be estimated.
We therefore assumed a uniform $T_{\rm ex}$ for such positions to derive $\tau{(^{13}\rm{CO})}$.
According to the $T_{\rm ex}$ map by \cite{Shimoikura2018} (see their Figure 6a), the maximum value of $T_{\rm ex}$ is 56 K in the vicinity of the H{$\,${\sc ii}} region. 
The values of $T_{\rm ex}$ exceeds 50 K only around the H{$\,${\sc ii}} region and decreases quickly toward the outside. 
The average $T_{\rm ex}$ over the W40 H{$\,${\sc ii}} region is about 20 K and that around Serpens South is about 11 K.
We calculated the total mass of the main component of the entire observed area to be $1.1\times10^{4}M_{\sun}$,
if we assume the uniform $T_{\rm ex}$ of 50 K throughout the observed area.
The ambiguity of $T_{\rm ex}$ does not affect the results of our analyses seriously, and the total mass would fluctuate only by $\sim5\%$ at most for a constant $T_{\rm ex}$ of $10-50$ K.



For the 40 km s$^{-1}$ component, the \tco emission line was not detected above the noise level at some positions.
Thus, we estimated the mass from the \co integrated intensity. 
For this, we measured the \co intensity in the range $36\leqq V_{\rm LSR}\leqq 46$  km s$^{-1}$ for the 40 km s$^{-1}$ component.
We used $1.8\times10^{20}$ cm$^{-2}$ K$^{-1}$/km s$^{-1}$ \citep{Dame2001}
to convert the \co integrated intensity to $N$(H$_2$). 
This yielded a mass estimate for this component of $150 M_{\sun}$.


\section{DISCUSSION} \label{sec:discussion}
The main component and the 40 km s$^{-1}$ component are widely distributed throughout the observed area. 
In this section, we discuss the relationship between these two components.

\subsection{Possible Interaction between the Two Components}

\begin{figure*}
\begin{center}
\includegraphics[scale=.4]{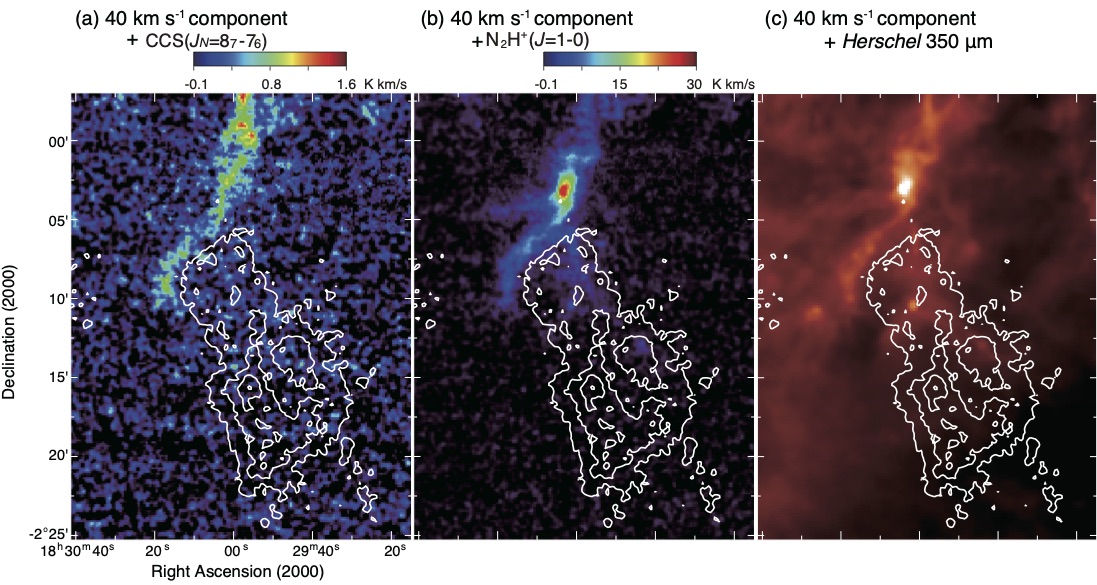}
\caption{
Comparison of the 40 km s$^{-1}$ component (contours, same as in Figure \ref{fig:spectra}) and other emission lines (color scale)
around Serpens South.
The 40 km s$^{-1}$ component overlaid with the 
(a) main component traced in the CCS emission line, 
(b) main component traced in the N$_2$H$^{+}$ emission line,
and (c) the $\it{Herschel}$ 350$\micron$ image.
The color scale image of the panels (a) and (b) are taken from \cite{Shimoikura2018}.
\label{fig:40}}
\end{center}
\end{figure*}

As shown in Figure \ref{fig:channel2}, we found that the spatial distributions of the two components appear to be anti-correlated with each other.
However, 
the main component traced in the \co emission is contaminated by heavy self-absorption
and we cannot identify the spatial structure clearly.
We therefore investigated the spatial relationship between the 40 km s$^{-1}$ component and the main component traced in other high density tracers such as N$_2$H$^{+}$ which we observed previously \citep{Shimoikura2018}. 
As a result, as shown in Figure \ref{fig:40}, 
we found that the dense filaments in Serpens South follow the edge of the 40 km s$^{-1}$ component with a clear spatial offset. 
In the region, the cloud consists of several filaments, as traced in the other emission lines,
and the young cluster is located where the filaments appear to intersect.
The 40 km s$^{-1}$ component lies at the edge of a filament extending to southeast from the young cluster.
As already mentioned, in the W40 region, the two components overlap in the vicinity of the cluster centered on IRS 5. 
These results strongly suggest that a collision between the main component and the 40 km s$^{-1}$ component occurred in W40 and Serpens South, which induced the cluster formation in both of the regions.

\begin{figure*}
\begin{center}
\includegraphics[scale=0.4]{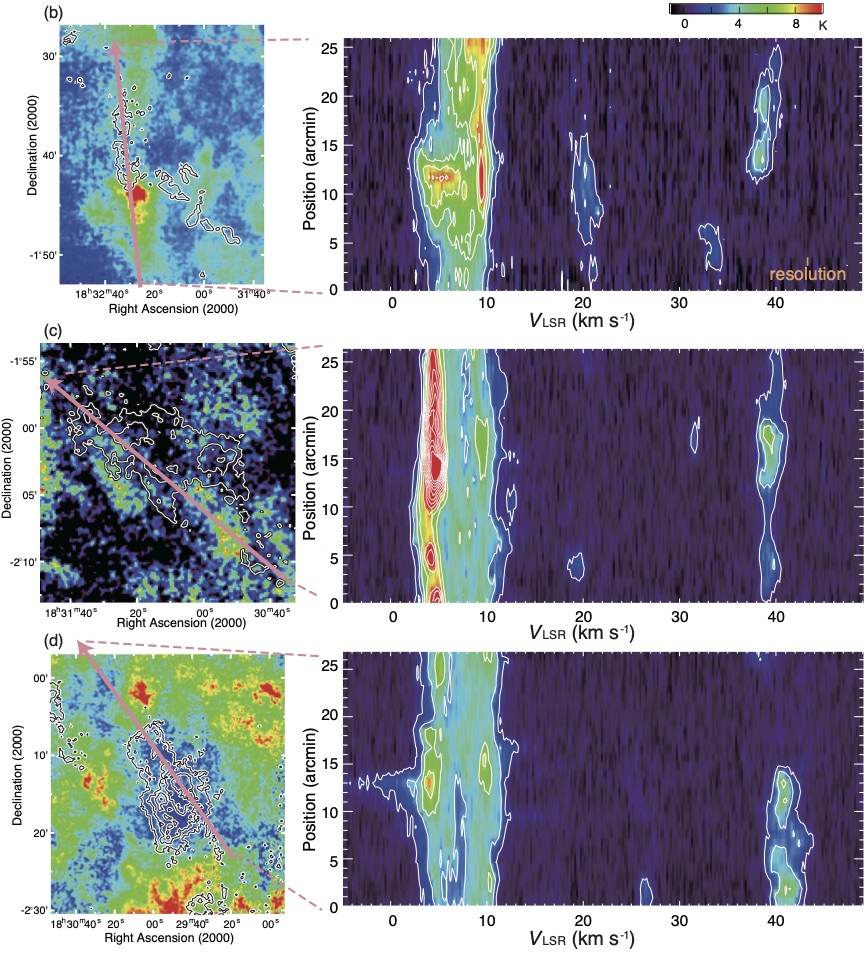}
\caption{
Position-velocity diagrams of the \co emission line taken along the cut in the left panels which are the same as in Figure \ref{fig:channel2} (b)--(d). 
The lowest contour and the contour interval of the position-velocity diagrams are 1.0 K and 2.0 K, respectively.
\label{fig:PV}}
\end{center}
\end{figure*}

In regions where the collision occurs, 
the numerical simulations of cloud-cloud collision demonstrated that in the early stage of the collision the interacting gas  dragged from the parent clouds tends to be distributed in between, having an intermediate velocity \citep{Habe1992,Haworth,Wu2017}.
In fact, such an intermediate-velocity feature has been suggested as a possible observational signature of a cloud collision and cloud interaction, 
and it has been observed in some star-forming regions \citep[e.g.,][]{Torii2015,dobashi2019b,dobashi2019a}.
In order to search for such a feature, we constructed position-velocity diagrams along the 40 km s$^{-1}$ as shown in Figure \ref{fig:PV}.
However, we did not detect such an intermediate-velocity feature above the noise level in this study.

\begin{figure*}
\begin{center}
\includegraphics[scale=0.4]{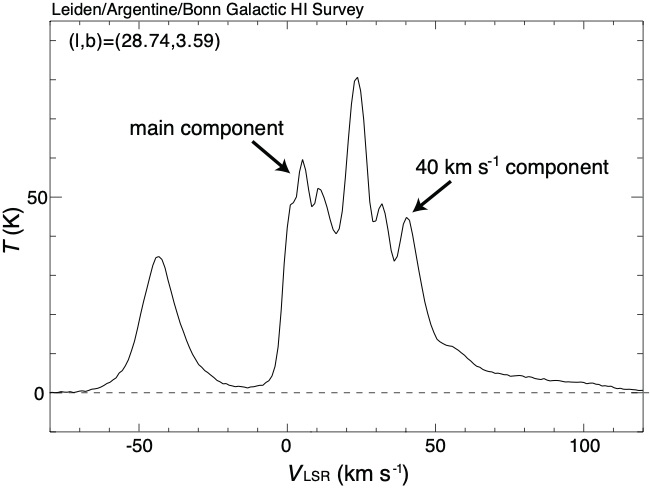}
\caption{
H{$\,${\sc i}} 21-cm spectrum toward the region of W40 and Serpens South taken from the LAB survey \citep{Kalberla2005}.  
\label{fig:HI}}
\end{center}
\end{figure*}

If the gas associated with an intermediate-velocity feature is not dense enough, however, 
molecular emission lines may not necessarily be excited, and 
it is also possible that dissociation of the molecular gas may occur.
Therefore, we investigated the velocity field of the atomic gas around the observed area using the Leiden/Argentine/Bonn (LAB) Survey of Galactic H{$\,${\sc i}} \citep{Kalberla2005}, a database of the 21-cm emission from neutral atomic hydrogen.
Figure \ref{fig:HI} shows the H{$\,${\sc i}} spectrum toward the observed area.
In addition to the main component and the 40 km s$^{-1}$ component, the other velocity components that we detected around $\simeq 25$ km s$^{-1}$ and $\simeq 35$ km s$^{-1}$  in the \co emission line are also detected in the 21-cm line.
However, as can be seen in the figure, because of the contamination by the other velocity components, 
we cannot identify the intermediate-velocity feature expected for the collision between the main component and the 40 km s$^{-1}$ component.

In order to constrain the distance to the 40 km s$^{-1}$ component, 
we calculated the kinematic distance using the
galactic rotation models from the literature \cite[e.g.,][]{Kerr, Wouterloot1990}. 
We used the velocity $V_{\rm LSR}=40$ km s$^{-1}$, $R_0$=8.5 kpc, and
$\Theta_0=220$ km s$^{-1}$, where $\Theta_0$ is the rotational speed of the Galaxy at $R_0$. 
We obtained the two kinematic distances of $\sim2.9$ kpc and $\sim12.0$ kpc, which correspond to the ``close" and ``far" distances in the model, respectively.
Since the galactic longitude of the observed region is $\ell \simeq28.8\arcdeg$, 
the 40 km s$^{-1}$ component would be located between the Scutum-Centaurus spiral Arm and the Sagittarius Arm 
in the former case, whereas it would be located in the Perseus Arm in the latter case.
In these cases, the mass of the 40 km s$^{-1}$ component that we obtained from the \co data assuming a distance of 436 pc
($150M_\sun$) must be rescaled to $6.8\times10^{3}M_\sun$ for $2.9$ kpc and to $2.3\times10^{5}M_\sun$ for $12.0$ kpc.

It is difficult to determine the distances to molecular clouds accurately and 
there is always a possibility that several molecular clouds lie by chance at different distances along the same line of sight.
However, the $^{12}$CO emission is about only 6 K in $T_{\rm mb}$ for the 40 km s$^{-1}$ component (see Figure \ref{fig:spectra}), 
so it is unlikely to be a massive molecular cloud with a mass of $6.8\times10^{3}M_\sun$ or more.
In addition, there are not many such massive molecular clouds in the sparse interstellar medium between spiral arms.
Therefore, it seems unlikely 
that the main and the 40 km s$^{-1}$ components are unrelated clouds lying on the same line of sight by chance.
In addition, 
the galactic latitude of W40 and Serpens South is $b\simeq3.5\arcdeg$ and
the two regions would be located 170 pc and 740 pc away from the Galactic plane 
if the distance were 2.8 kpc and 12.1 kpc, respectively.
Considering that the scale height of the galactic disk is $\sim100$ pc, these are not realistic solutions.
We conclude that the 40 km s$^{-1}$ component is likely to be a cloud located at the same distance as the main component, 
and that they are physically interacting with each other. 

\subsection{Origin of the 40 km s$^{-1}$ Component}

We consider the origin of the 40 km s$^{-1}$ component. 
According to \cite{deGeus1992}, \cite{Frisch1995}, and \cite{Breitschwerdt1996}, 
the Aquila Rift where W40 and Serpens South are located, may be influenced by the expanding gas superbubble shells created by star formation in the Scorpius-Centaurus Association (Sco OB2). 
Sco OB2 is composed of three subgroups, 
i.e., Upper Centaurus-Lupus (aged {\color{red}$\sim$}17 Myr), Lower Centaurus-Crux (aged {\color{red}$\sim$}16 Myr), and Upper Scorpius (aged {\color{red}$\sim$}5 Myr) \citep[e.g.,][]{Preibisch2008, Krause}.  
Several superbubbles have been observed around these subgroups, which are H{$\,${\sc i}} shells formed by stellar winds and supernova explosions of stars in Sco OB2 \cite[e.g.,][]{deGeus1992,Maiz2001}.
\cite{Nakamura2017} found that the molecular gas distribution around W40 and Serpens South is complicated, and they suggested that this may be due to the interaction of the molecular clouds with the superbubbles.
We suggest that the 40 km s$^{-1}$ component also may be molecular gas influenced by the superbubbles created by Sco OB2.

The stellar winds of the supperbubbles from Sco OB2 are mainly divided into the following three shells, which formed during three epochs of
star formation over the past 15 Myrs \citep[e.g.,][]{deGeus1992,Preibisch2008}:
\begin{enumerate}
\renewcommand{\labelenumi}{(\Alph{enumi})}
\item the outermost boundary of the Sco OB2 superbubble shell,
\item the shell that seems to be caused by star formation in Upper Scorpius, and
\item the Loop I supernova remnant. 
\end{enumerate}

\begin{figure*}
\begin{center}
\includegraphics[scale=.4]{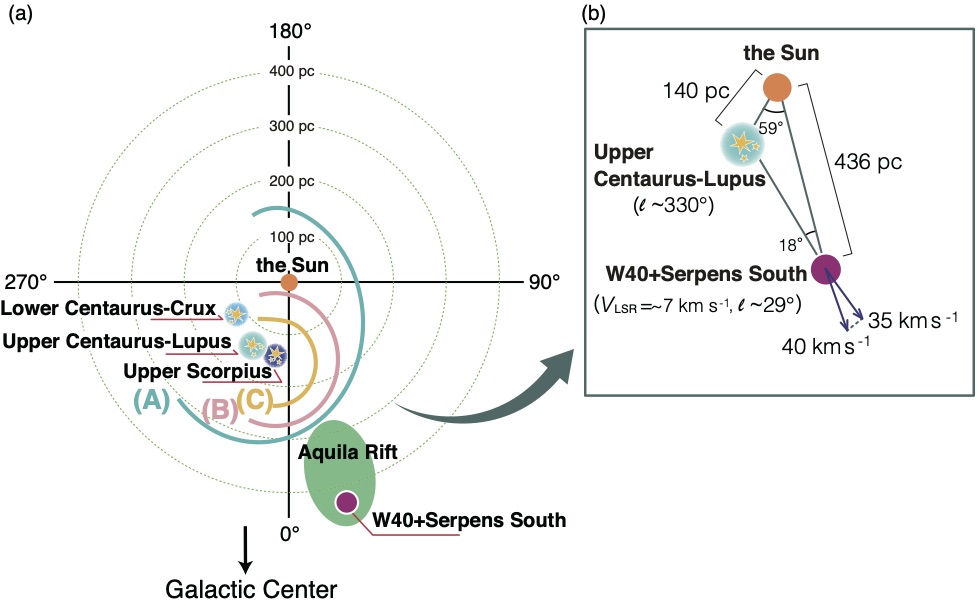}
\caption{
(a) Schematic illustration of W40, Serpens South, and the Scorpius-Centaurus Association (Sco OB2),
which is constructed from the summary by \cite{Breitschwerdt1996}.
Sco OB2 is composed of three subgroups, 
i.e., Upper Centaurus-Lupus, Lower Centaurus-Crux, and Upper Scorpius \citep[e.g.,][]{Preibisch2008}. 
(b)  A relationship diagram among W40+Serpens South, Upper Centaurus-Lupus, and the Sun.
\label{fig:model}}
\end{center}
\end{figure*}

The dynamical ages of these bubbles are estimated to be 15 Myr, 4 Myr, and 0.25 Myr, respectively \citep{deGeus1992, Frisch1995, Breitschwerdt1996}.
Figure \ref{fig:model} summarizes the positional relationships of the Sun, W40+Serpens South, and Sco OB2. 
The shells (A) -- (C) in the above correspond to (A) -- (C) in Figure \ref{fig:model}. 

Assuming that the 40 km s$^{-1}$ component was swept up to its current position by stellar winds from Sco OB2,
we attempt to estimate which shell among (A) -- (C) is the origin of the component.
If we assume that the distance between the Sun and W40+Serpens South ($\ell \simeq29\arcdeg$) is 436 pc \citep{Ortiz2018}, whereas the distance from the Sun to Sco OB2 ($\ell \simeq330\arcdeg$), which is the center of  Upper Centaurus-Lupus, is 140 pc \citep{de_Zeeuw1999}, 
the distance between W40+Serpens South and Upper Centaurus-Lupus is about 380 pc.
If the molecular gas of the 40 km s$^{-1}$ component originating from Sco OB2 
and was accelerated by the stellar wind in the direction of W40+Serpens South, 
the relative velocity is 40/cos($18\arcdeg)-7=35$ km s$^{-1}$. 
Because the main component has at $V_{\rm LSR}\simeq 7$ km s$^{-1}$,
the time required for the movement is about 11 Myr.
This is similar to the dynamical age of (A) (15 Myr) in the above, 
suggesting that the 40 km s$^{-1}$ component is the gas swept up by the shell (A).

\cite{Maiz2001} investigated the proper motion of the OB stars in the three subgroups and calculated the positions of the subgroups in the past.
Their results show that each subgroup was closer to the present Sun position $\sim5$ Myr ago,
suggesting a possibility that the 40 km s$^{-1}$ component could be the gas swept up by
the shell (B) rather than (A).
Recently, \cite{Krause} analyzed observational data of gas+dust and performed 3D hydrodynamical simulations for Sco OB2.
Their simulations showed that the expansion of the superbubble shell (= shell A) initiated about 15 Myr ago in Upper Centaurus-Lupus and caused a sequence of star formation in Sco OB2, which is a consistent picture with what can be observed today.
In this paper, we suggest that the 40 km s$^{-1}$ component was caused by the shell (A) because of the rough match in the dynamical age and the movement time as well as because of the results of the simulations, although there are large uncertainties such as past and current three-dimensional motions of the Aquila Rift complex.


If a molecular cloud is compressed by the collision, the free-fall time is shortened due to the compression. 
Since the timescale of star formation is about the free-fall time, the gas compression by the collisions is expected to induce star formation. 
The age of the YSOs in W40 is about 1 Myr \citep{Kuhn}, 
and so the collision between the main component and the 40 km s$^{-1}$ component may have occurred about 1 Myr ago.

In summary, we suggest that the 40 km s$^{-1}$ component is the gas swept up by the stellar wind (A) from Sco OB2, which caused a collision with W40 and Serpens South about 1 Myr ago.
This may have compressed the molecular clouds and triggered star formation in W40 and Serpens South.
Later, the expansion of the H{$\,${\sc ii}} region in W40 further induced formation of the young cluster in Serpens South, as suggested by \cite{Shimoikura2018}.


\section{Conclusion}

We have studied the active star-forming regions W40 and Serpens South in the Aquila Rift complex.
We observed a region of 1 square degree around W40 and Serpens South 
using the 45 m radio telescope at NRO
and examined the distributions and velocity structures of the molecular gas using mainly the $^{12}$CO and $^{13}$CO molecular lines.
We obtained the following results.

\begin{enumerate}
\item The molecular clouds in W40 and Serpens South have radial velocities mostly around $V_{\rm LSR} \simeq 7$ km s$^{-1}$, which we call the ``main component".
In addition, we found three different components in the observed area at 
$V_{\rm LSR}\simeq 20, 25, 35$, and 40 km s$^{-1}$.
In particular, there exists a significant amount of molecular gas at $V_{\rm LSR}\simeq 40$ km s$^{-1}$.
This component is widely distributed across the observed area, 
and we call it the ``40 km s$^{-1}$ component".

\item Assuming local thermodynamic equilibrium, 
the mass of the main component calculated from the $^{13}$CO emission line
is estimated to be $1.1\times10^{4}M_{\sun}$.
For the 40 km s$^{-1}$ component, 
we calculate the mass from the integrated intensity of the $^{12}$CO line, 
assuming the conversion factor given by \cite{Dame2001}, to be about $150 M_{\sun}$ at a distance of 436 pc.

\item On the basis of the distributions of the $^{12}$CO and $^{13}$CO lines on the plane of the sky,
we investigated the relationship between the main component and the 40 km s$^{-1}$ component.
We found that there is an anti-correlation between the two components
and that they overlap in the vicinity of a cluster centered on the B-type star IRS 5 in a compact H{$\,${\sc ii}} region. 
The 40 km s$^{-1}$ component is also distributed along a filament in the vicinity of the young cluster in Serpens South.
From these results, we suggest that the 40 km s$^{-1}$ component interacting with the main component
and induced star formation in W40 and Serpens South. 

\item The presence of bubble-like winds around the two regions induced by star formation activity in the Scorpius-Centaurus Association (Sco OB2) has been reported in the literature.
We discussed the possible origin of the 40 km s$^{-1}$ component in terms of gas swept up by the stellar winds from Sco OB2. We suggest that the 40 km s$^{-1}$ component collided with W40 and Serpens South about 1 Myr ago, triggering star formation in W40 and Serpens South.
\end{enumerate}


\acknowledgments
We are very grateful to the anonymous referee for providing
useful comments and suggestions to improve this paper.
We thank the other members of Star Formation Legacy project for their support during the observations.
This work was supported by JSPS KAKENHI Grant Numbers 17K00963, 17H02863, 17H01118, 19H05070. 
The 45 m radio telescope is operated by NRO, a branch of NAOJ. 



\begin{thebibliography}{}
\expandafter\ifx\csname natexlab\endcsname\relax\def\natexlab#1{#1}\fi
\providecommand{\url}[1]{\href{#1}{#1}}
\providecommand{\dodoi}[1]{doi:~\href{http://doi.org/#1}{\nolinkurl{#1}}}
\providecommand{\doeprint}[1]{\href{http://ascl.net/#1}{\nolinkurl{http://ascl.net/#1}}}
\providecommand{\doarXiv}[1]{\href{https://arxiv.org/abs/#1}{\nolinkurl{https://arxiv.org/abs/#1}}}

\bibitem[{{Andr{\'e}} {et~al.}(2010){Andr{\'e}}, {Men'shchikov}, {Bontemps},
  {K{\"o}nyves}, {Motte}, {Schneider}, {Didelon}, {Minier}, {Saraceno},
  {Ward-Thompson}, {di Francesco}, {White}, {Molinari}, {Testi}, {Abergel},
  {Griffin}, {Henning}, {Royer}, {Mer{\'{\i}}n}, {Vavrek}, {Attard},
  {Arzoumanian}, {Wilson}, {Ade}, {Aussel}, {Baluteau}, {Benedettini},
  {Bernard}, {Blommaert}, {Cambr{\'e}sy}, {Cox}, {di Giorgio}, {Hargrave},
  {Hennemann}, {Huang}, {Kirk}, {Krause}, {Launhardt}, {Leeks}, {Le Pennec},
  {Li}, {Martin}, {Maury}, {Olofsson}, {Omont}, {Peretto}, {Pezzuto}, {Prusti},
  {Roussel}, {Russeil}, {Sauvage}, {Sibthorpe}, {Sicilia-Aguilar}, {Spinoglio},
  {Waelkens}, {Woodcraft}, \& {Zavagno}}]{Andre2010}
{Andr{\'e}}, P., {Men'shchikov}, A., {Bontemps}, S., {et~al.} 2010, \aap, 518,
  L102, \dodoi{10.1051/0004-6361/201014666}

\bibitem[{Breitschwerdt {et~al.}(1996)Breitschwerdt, Egger, Freyberg, Frisch,
  \& Vallerga}]{Breitschwerdt1996}
Breitschwerdt, D., Egger, R., Freyberg, M., Frisch, P., \& Vallerga, J. 1996,
  Space Science Reviews, 78, 183, \dodoi{10.1007/bf00170805}

\bibitem[{Dame {et~al.}(2001)Dame, Hartmann, \& Thaddeus}]{Dame2001}
Dame, T.~M., Hartmann, D., \& Thaddeus, P. 2001, The Astrophysical Journal,
  547, 792, \dodoi{10.1086/318388}

\bibitem[{{de Geus}(1992)}]{deGeus1992}
{de Geus}, E.~J. 1992, \aap, 262, 258

\bibitem[{de~Zeeuw {et~al.}(1999)de~Zeeuw, Hoogerwerf, de~Bruijne, Brown, \&
  Blaauw}]{de_Zeeuw1999}
de~Zeeuw, P.~T., Hoogerwerf, R., de~Bruijne, J. H.~J., Brown, A. G.~A., \&
  Blaauw, A. 1999, The Astronomical Journal, 117, 354, \dodoi{10.1086/300682}

\bibitem[{{Dickman}(1978)}]{Dickman1978}
{Dickman}, R.~L. 1978, \apjs, 37, 407, \dodoi{10.1086/190535}

\bibitem[{{Dobashi}(2011)}]{Dobashi2011}
{Dobashi}, K. 2011, \pasj, 63, 1, \dodoi{10.1093/pasj/63.sp1.S1}

\bibitem[{{Dobashi} {et~al.}(2019{\natexlab{a}}){Dobashi}, {Shimoikura},
  {Endo}, {Takagi}, {Nakamura}, {Shimajiri}, \& {Bernard}}]{dobashi2019b}
{Dobashi}, K., {Shimoikura}, T., {Endo}, N., {et~al.} 2019{\natexlab{a}},
  \pasj, 71, S11, \dodoi{10.1093/pasj/psy122}

\bibitem[{{Dobashi} {et~al.}(2019{\natexlab{b}}){Dobashi}, {Shimoikura},
  {Katakura}, {Nakamura}, \& {Shimajiri}}]{dobashi2019a}
{Dobashi}, K., {Shimoikura}, T., {Katakura}, S., {Nakamura}, F., \&
  {Shimajiri}, Y. 2019{\natexlab{b}}, \pasj, 71, S12,
  \dodoi{10.1093/pasj/psz041}

\bibitem[{{Dobashi} {et~al.}(2005){Dobashi}, {Uehara}, {Kandori}, {Sakurai},
  {Kaiden}, {Umemoto}, \& {Sato}}]{Dobashi2005}
{Dobashi}, K., {Uehara}, H., {Kandori}, R., {et~al.} 2005, \pasj, 57, 1

\bibitem[{{Frisch} \& {Dwarkadas}(2018)}]{Frisch2018}
{Frisch}, P., \& {Dwarkadas}, V.~V. 2018, arXiv e-prints, arXiv:1801.06223.
\newblock \doarXiv{1801.06223}

\bibitem[{{Frisch}(1995)}]{Frisch1995}
{Frisch}, P.~C. 1995, \ssr, 72, 499, \dodoi{10.1007/BF00749006}

\bibitem[{{Gutermuth} {et~al.}(2008){Gutermuth}, {Bourke}, {Allen}, {Myers},
  {Megeath}, {Matthews}, {J{\o}rgensen}, {Di Francesco}, {Ward-Thompson},
  {Huard}, {Brooke}, {Dunham}, {Cieza}, {Harvey}, \& {Chapman}}]{Gutermuth2008}
{Gutermuth}, R.~A., {Bourke}, T.~L., {Allen}, L.~E., {et~al.} 2008, \apjl, 673,
  L151, \dodoi{10.1086/528710}

\bibitem[{{Habe} \& {Ohta}(1992)}]{Habe1992}
{Habe}, A., \& {Ohta}, K. 1992, \pasj, 44, 203

\bibitem[{{Haworth} {et~al.}(2015){Haworth}, {Tasker}, {Fukui}, {Torii},
  {Dale}, {Shima}, {Takahira}, {Habe}, \& {Hasegawa}}]{Haworth}
{Haworth}, T.~J., {Tasker}, E.~J., {Fukui}, Y., {et~al.} 2015, \mnras, 450, 10,
  \dodoi{10.1093/mnras/stv639}

\bibitem[{{Kalberla, P. M. W.} {et~al.}(2005){Kalberla, P. M. W.}, {Burton, W.
  B.}, {Hartmann, Dap}, {Arnal, E. M.}, {Bajaja, E.}, {Morras, R.}, \&
  {P\"oppel, W. G. L.}}]{Kalberla2005}
{Kalberla, P. M. W.}, {Burton, W. B.}, {Hartmann, Dap}, {et~al.} 2005, A\&A,
  440, 775, \dodoi{10.1051/0004-6361:20041864}

\bibitem[{{Kerr} \& {Lynden-Bell}(1986)}]{Kerr}
{Kerr}, F.~J., \& {Lynden-Bell}, D. 1986, \mnras, 221, 1023

\bibitem[{{Koenig} {et~al.}(2012){Koenig}, {Leisawitz}, {Benford}, {Rebull},
  {Padgett}, \& {Assef}}]{Koenig}
{Koenig}, X.~P., {Leisawitz}, D.~T., {Benford}, D.~J., {et~al.} 2012, \apj,
  744, 130, \dodoi{10.1088/0004-637X/744/2/130}

\bibitem[{{K{\"o}nyves} {et~al.}(2015){K{\"o}nyves}, {Andr{\'e}},
  {Men'shchikov}, {Palmeirim}, {Arzoumanian}, {Schneider}, {Roy}, {Didelon},
  {Maury}, {Shimajiri}, {Di Francesco}, {Bontemps}, {Peretto}, {Benedettini},
  {Bernard}, {Elia}, {Griffin}, {Hill}, {Kirk}, {Ladjelate}, {Marsh}, {Martin},
  {Motte}, {Nguy{\^e}n Luong}, {Pezzuto}, {Roussel}, {Rygl}, {Sadavoy},
  {Schisano}, {Spinoglio}, {Ward-Thompson}, \& {White}}]{Konyves2015}
{K{\"o}nyves}, V., {Andr{\'e}}, P., {Men'shchikov}, A., {et~al.} 2015, \aap,
  584, A91, \dodoi{10.1051/0004-6361/201525861}

\bibitem[{{Krause} {et~al.}(2018){Krause}, {Burkert}, {Diehl}, {Fierlinger},
  {Gaczkowski}, {Kroell}, {Ngoumou}, {Roccatagliata}, {Siegert}, \&
  {Preibisch}}]{Krause}
{Krause}, M. G.~H., {Burkert}, A., {Diehl}, R., {et~al.} 2018, \aap, 619, A120,
  \dodoi{10.1051/0004-6361/201732416}

\bibitem[{{Kuhn} {et~al.}(2010){Kuhn}, {Getman}, {Feigelson}, {Reipurth},
  {Rodney}, \& {Garmire}}]{Kuhn}
{Kuhn}, M.~A., {Getman}, K.~V., {Feigelson}, E.~D., {et~al.} 2010, \apj, 725,
  2485, \dodoi{10.1088/0004-637X/725/2/2485}

\bibitem[{{Kusune} {et~al.}(2019){Kusune}, {Nakamura}, {Sugitani}, {Sato},
  {Tamura}, {Kwon}, {Dobashi}, {Shimoikura}, \& {Wu}}]{Kusune2019}
{Kusune}, T., {Nakamura}, F., {Sugitani}, K., {et~al.} 2019, \pasj, 71, S5,
  \dodoi{10.1093/pasj/psz040}

\bibitem[{Ma{\'\i}z-Apell{\'a}niz(2001)}]{Maiz2001}
Ma{\'\i}z-Apell{\'a}niz, J. 2001, The Astrophysical Journal, 560, L83,
  \dodoi{10.1086/324016}

\bibitem[{{Mallick} {et~al.}(2013){Mallick}, {Kumar}, {Ojha}, {Bachiller},
  {Samal}, \& {Pirogov}}]{Mallick}
{Mallick}, K.~K., {Kumar}, M.~S.~N., {Ojha}, D.~K., {et~al.} 2013, \apj, 779,
  113, \dodoi{10.1088/0004-637X/779/2/113}

\bibitem[{{Maury} {et~al.}(2011){Maury}, {Andr{\'e}}, {Men'shchikov},
  {K{\"o}nyves}, \& {Bontemps}}]{Maury2011}
{Maury}, A.~J., {Andr{\'e}}, P., {Men'shchikov}, A., {K{\"o}nyves}, V., \&
  {Bontemps}, S. 2011, \aap, 535, A77, \dodoi{10.1051/0004-6361/201117132}

\bibitem[{{Minamidani} {et~al.}(2016){Minamidani}, {Nishimura}, {Miyamoto},
  {Kaneko}, {Iwashita}, {Miyazawa}, {Nishitani}, {Wada}, {Fujii}, {Takahashi},
  {Iizuka}, {Ogawa}, {Kimura}, {Kozuki}, {Hasegawa}, {Matsuo}, {Fujita},
  {Ohashi}, {Morokuma-Matsui}, {Maekawa}, {Muraoka}, {Nakajima}, {Umemoto},
  {Sorai}, {Nakamura}, {Kuno}, \& {Saito}}]{Minamidani}
{Minamidani}, T., {Nishimura}, A., {Miyamoto}, Y., {et~al.} 2016, in \procspie,
  Vol. 9914, Millimeter, Submillimeter, and Far-Infrared Detectors and
  Instrumentation for Astronomy VIII, 99141Z, \dodoi{10.1117/12.2232137}

\bibitem[{{Nakamura} {et~al.}(2017){Nakamura}, {Dobashi}, {Shimoikura},
  {Tanaka}, \& {Onishi}}]{Nakamura2017}
{Nakamura}, F., {Dobashi}, K., {Shimoikura}, T., {Tanaka}, T., \& {Onishi}, T.
  2017, \apj, 837, 154, \dodoi{10.3847/1538-4357/aa5ea6}

\bibitem[{Nakamura {et~al.}(2011)Nakamura, Sugitani, Shimajiri, Tsukagoshi,
  Higuchi, Nishiyama, Kawabe, Takami, Karr, Gutermuth, \&
  et~al.}]{Nakamura2011}
Nakamura, F., Sugitani, K., Shimajiri, Y., {et~al.} 2011, The Astrophysical
  Journal, 737, 56, \dodoi{10.1088/0004-637x/737/2/56}

\bibitem[{{Nakamura} {et~al.}(2014){Nakamura}, {Sugitani}, {Tanaka},
  {Nishitani}, {Dobashi}, {Shimoikura}, {Shimajiri}, {Kawabe}, {Yonekura},
  {Mizuno}, {Kimura}, {Tokuda}, {Kozu}, {Okada}, {Hasegawa}, {Ogawa}, {Kameno},
  {Shinnaga}, {Momose}, {Nakajima}, {Onishi}, {Maezawa}, {Hirota}, {Takano},
  {Iono}, {Kuno}, \& {Yamamoto}}]{Nakamura2014}
{Nakamura}, F., {Sugitani}, K., {Tanaka}, T., {et~al.} 2014, \apjl, 791, L23,
  \dodoi{10.1088/2041-8205/791/2/L23}

\bibitem[{{Nakamura} {et~al.}(2019){Nakamura}, {Ishii}, {Dobashi},
  {Shimoikura}, {Shimajiri}, {Kawabe}, {Tanabe}, {Hirose}, {Oyamada},
  {Urasawa}, {Takemura}, {Tsukagoshi}, {Momose}, {Sugitani}, {Nishi},
  {Okumura}, {Sanhueza}, {Nygen-Luong}, \& {Kusune}}]{Nakamura2019}
{Nakamura}, F., {Ishii}, S., {Dobashi}, K., {et~al.} 2019, arXiv e-prints,
  arXiv:1909.05980.
\newblock \doarXiv{1909.05980}

\bibitem[{Ortiz-Le{\'o}n {et~al.}(2018)Ortiz-Le{\'o}n, Loinard, Dzib, Kounkel,
  Galli, Tobin, II, Hartmann, Rodr{\'\i}guez, Brice{\~n}o, \&
  et~al.}]{Ortiz2018}
Ortiz-Le{\'o}n, G.~N., Loinard, L., Dzib, S.~A., {et~al.} 2018, The
  Astrophysical Journal, 869, L33, \dodoi{10.3847/2041-8213/aaf6ad}

\bibitem[{{Pirogov} {et~al.}(2013){Pirogov}, {Ojha}, {Thomasson}, {Wu}, \&
  {Zinchenko}}]{Pirogov2013}
{Pirogov}, L., {Ojha}, D.~K., {Thomasson}, M., {Wu}, Y.-F., \& {Zinchenko}, I.
  2013, \mnras, 436, 3186, \dodoi{10.1093/mnras/stt1802}

\bibitem[{{Preibisch} \& {Mamajek}(2008)}]{Preibisch2008}
{Preibisch}, T., \& {Mamajek}, E. 2008, {The Nearest OB Association:
  Scorpius-Centaurus (Sco OB2)}, ed. B.~{Reipurth}, Vol.~5, 235

\bibitem[{{Rodney} \& {Reipurth}(2008)}]{Rodney}
{Rodney}, S.~A., \& {Reipurth}, B. 2008, {The W40 Cloud Complex}, ed.
  B.~{Reipurth}, 683

\bibitem[{{Sawada} {et~al.}(2008){Sawada}, {Ikeda}, {Sunada}, {Kuno},
  {Kamazaki}, {Morita}, {Kurono}, {Koura}, {Abe}, {Kawase}, {Maekawa},
  {Horigome}, \& {Yanagisawa}}]{Sawada}
{Sawada}, T., {Ikeda}, N., {Sunada}, K., {et~al.} 2008, \pasj, 60, 445,
  \dodoi{10.1093/pasj/60.3.445}

\bibitem[{{Sharpless}(1959)}]{Sharpless}
{Sharpless}, S. 1959, \apjs, 4, 257, \dodoi{10.1086/190049}

\bibitem[{{Shimoikura} {et~al.}(2015){Shimoikura}, {Dobashi}, {Nakamura},
  {Hara}, {Tanaka}, {Shimajiri}, {Sugitani}, \& {Kawabe}}]{Shimoikura2015}
{Shimoikura}, T., {Dobashi}, K., {Nakamura}, F., {et~al.} 2015, \apj, 806, 201,
  \dodoi{10.1088/0004-637X/806/2/201}

\bibitem[{Shimoikura {et~al.}(2018)Shimoikura, Dobashi, Nakamura, Shimajiri, \&
  Sugitani}]{Shimoikura2018}
Shimoikura, T., Dobashi, K., Nakamura, F., Shimajiri, Y., \& Sugitani, K. 2018,
  Publications of the Astronomical Society of Japan,
  \dodoi{10.1093/pasj/psy115}

\bibitem[{{Shuping} {et~al.}(2012){Shuping}, {Vacca}, {Kassis}, \&
  {Yu}}]{Shuping2012}
{Shuping}, R.~Y., {Vacca}, W.~D., {Kassis}, M., \& {Yu}, K.~C. 2012, \aj, 144,
  116, \dodoi{10.1088/0004-6256/144/4/116}

\bibitem[{{Smith} {et~al.}(1985){Smith}, {Bentley}, {Castelaz}, {Gehrz},
  {Grasdalen}, \& {Hackwell}}]{Smith}
{Smith}, J., {Bentley}, A., {Castelaz}, M., {et~al.} 1985, \apj, 291, 571,
  \dodoi{10.1086/163097}

\bibitem[{{Sugitani} {et~al.}(2010){Sugitani}, {Nakamura}, {Tamura},
  {Watanabe}, {Kandori}, {Nishiyama}, {Kusakabe}, {Hashimoto}, {Nagata}, \&
  {Sato}}]{Sugitani2010}
{Sugitani}, K., {Nakamura}, F., {Tamura}, M., {et~al.} 2010, \apj, 716, 299,
  \dodoi{10.1088/0004-637X/716/1/299}

\bibitem[{{Torii} {et~al.}(2015){Torii}, {Hasegawa}, {Hattori}, {Sano},
  {Ohama}, {Yamamoto}, {Tachihara}, {Soga}, {Shimizu}, {Okuda}, {Mizuno},
  {Onishi}, {Mizuno}, \& {Fukui}}]{Torii2015}
{Torii}, K., {Hasegawa}, K., {Hattori}, Y., {et~al.} 2015, \apj, 806, 7,
  \dodoi{10.1088/0004-637X/806/1/7}

\bibitem[{{Vallee}(1987)}]{Vallee}
{Vallee}, J.~P. 1987, \aap, 178, 237

\bibitem[{{Wouterloot} {et~al.}(1990){Wouterloot}, {Brand}, {Burton}, \&
  {Kwee}}]{Wouterloot1990}
{Wouterloot}, J.~G.~A., {Brand}, J., {Burton}, W.~B., \& {Kwee}, K.~K. 1990,
  \aap, 230, 21

\bibitem[{{Wu} {et~al.}(2017){Wu}, {Tan}, {Christie}, {Nakamura}, {Van Loo}, \&
  {Collins}}]{Wu2017}
{Wu}, B., {Tan}, J.~C., {Christie}, D., {et~al.} 2017, \apj, 841, 88,
  \dodoi{10.3847/1538-4357/aa6ffa}

\bibitem[{{Zhu} {et~al.}(2006){Zhu}, {Wu}, \& {Wei}}]{Zhu}
{Zhu}, L., {Wu}, Y.-F., \& {Wei}, Y. 2006, \cjaa, 6, 61,
  \dodoi{10.1088/1009-9271/6/1/007}

\end{thebibliography}

\clearpage



\clearpage

\end{document}